\documentclass[reprint,amsmath,amssymb,aps]{revtex4-2}

\usepackage{amssymb}
\usepackage{graphicx}
\usepackage{dcolumn}
\usepackage{bm}
\usepackage{textcomp}
\usepackage{amsmath}
\usepackage{hyperref}
\usepackage{ntheorem}
\usepackage{verbatim}
\usepackage{cleveref}
\usepackage[titletoc]{appendix}
\newcommand\ket[1]{\ensuremath{|#1\rangle}}

\newcommand\mean[1]{\ensuremath{\langle #1\rangle}}
\newtheorem{theorem}{Theorem}
\newtheorem*{proof}{Proof}
\newtheorem{remark}{Remark}

\begin{document}
\title {Sending-or-Not-Sending Twin-Field Quantum Key Distribution with Redundant Space}

\author{ Hai Xu$^{1,2}$,  Xiao-Long Hu$ ^{1,3}$, Cong Jiang$ ^{1,4}$, Zong-Wen Yu$ ^{1,5}$, 
and Xiang-Bin Wang$ ^{1,2,4,6,7\footnote{Email Address: xbwang@mail.tsinghua.edu.cn}\footnote{Also at Center for Atomic and Molecular Nanosciences, Tsinghua University, Beijing 100084, China}}$}
\affiliation{ \centerline{$^{1}$State Key Laboratory of Low Dimensional Quantum Physics, Department of Physics,}
\centerline{Tsinghua University, Beijing 100084, China}
\centerline{$^{2}$ Frontier Science Center for Quantum Information, Beijing, China}
\centerline{$^{3}$  School of Physics, State Key Laboratory of Optoelectronic Materials and Technologies,}
\centerline{Sun Yat-sen University, Guangzhou 510275, China}
\centerline{$^{4}$ Jinan Institute of Quantum technology, SAICT, Jinan 250101, China}
\centerline{$^{5}$Data Communication Science and Technology Research Institute, Beijing 100191, China}
\centerline{$^{6}$ Shanghai Branch, CAS Center for Excellence and Synergetic Innovation Center in Quantum Information and Quantum Physics,}
\centerline{University of Science and Technology of China, Shanghai 201315, People’s Republic of China}
\centerline{$^{7}$ Shenzhen Institute for Quantum Science and Engineering, and Physics Department,}
\centerline{Southern University of Science and Technology, Shenzhen 518055, China}}

\begin{abstract}
We propose to adopt redundant space such as polarization mode in the sending-or-not-sending Twin-Field quantum key distribution (TF-QKD) in the Fock space. With the help of redundant space such as photon polarization, we can post-select events according to the outcome of the observation to the additional quantity. This compresses the bit-flip error rate in the post-selected events of the SNS protocol. The calculation shows that the method using redundant space can greatly improve the performance in practical  TF-QKD, especially when the total number of pulses is small. 
\end{abstract}

\maketitle
\section {Introduction}
Quantum Key Distribution (QKD) provides a promising approach to unconditionally secure communication~\cite{bennett2014quantum,Pirandola2020advance,xu2020secure,gisin2002quantum,scarani2009security,hwang2003quantum,wang2005beating,lo2005decoy,lo2012measurement,braunstein2012side}. The decoy-state method~\cite{hwang2003quantum,wang2005beating,lo2005decoy} can keep the unconditional security of QKD when the imperfect single photon sources are used, and those security loopholes caused by imperfect detection devices can be closed by Measurement-Device-Independent (MDI)-QKD~\cite{braunstein2012side,lo2012measurement,wang2013three}. With the development of both theories and experiments in recent years~\cite{rubenok2013real,wang2015phase,comandar2016quantum,yin2016measurement,boaron2018secure,lucamarini2018overcoming,curty2014finite,muller2009composability,renner2005security,konig2007small,tomamichel2012tight,pirandola2015high,wang2017measurement,semenenko2020chip,cao2020chip,cao2020long,xu2013practical,xu2014protocol,yu2015statistical,zhou2016making,liu2019reference,hu2021practical,jiang2021higher,teng2021sending,Fan2021measurement,Fan2022robust,liao2017satellite,chen2021integrated},
QKD is gradually maturing to real world deployment over optical fiber and satellite ~\cite{liao2017satellite,Pirandola2020advance,xu2020secure,chen2021integrated,lu2022micius}.
Especially, the emergence of Twin-Field (TF)-QKD~\cite{lucamarini2018overcoming}  and its variants~\cite{wang2018twin,tamaki2018information,ma2018phase,lin2018simple,cui2019twin,curty2019simple}  have improved the secure distance drastically both in laboratory optical fiber experiments ~\cite{minder2019experimental,fang2020implementation,liu2019experimental,wang2019beating,pittaluga2021600,wang2022twin,chen2020sending,chen2022quantum} and field test~\cite{chen2021twin,liu2021field,clivati2022coherent}.
 Note that, TF-QKD is also free from detection loopholes. 

 It is necessary to generate fresh secure keys for instant use in many practical application scenarios.
 However, when the finite size effect and statistical fluctuations~\cite{muller2009composability,renner2005security,konig2007small,tomamichel2012tight,curty2014finite,lim2014concise,curras2021tight,maeda2019repeaterless,yin2019finite,jiang2019unconditional,jiang2020zigzag,jiang2021composable,chen2020sending,chen2021twin,liu2021field,clivati2022coherent,chen2022quantum,wang2022twin} are taken into consideration, one needs to take hours for a QKD protocol to accumulate a sufficient amount of data, when the number of total pulses is assumed to be larger than $10^{12}$~\cite{yin2016measurement,chen2022quantum,chen2021twin,wang2022twin}. 
It seems to be rather challenging to generate considerable secure keys with a small data size, such as $10^8\sim 10^{10}$, which can be completed in a few seconds, given the system repetition frequency of $100$ MHz$ \sim 1$ GHz. 

The sending-or-not-sending (SNS) TF-QKD protocol~\cite{wang2018twin,yu2019sending,hu2019sending,jiang2019unconditional,xu2020sending,jiang2020zigzag,jiang2021composable,jiang2022robust,hu2022universal,hu2022sending} does not take post-selection for the phase of signal pulses, hence the key generation becomes more efficient, and the traditional decoy-state analysis and the existing theories of finite-key effects can directly apply.
What's more important, the SNS protocol does not request single-photon interference for signal pulses, which makes it robust to misalignment errors.
 However, the original SNS protocol uses a small sending probability to control the bit-flip error rate. Through two-way classical communication~\cite{chau2002practical,gottesman2003proof,wang2004quantum,kraus2007security,bae2007key}, one can take error rejection by parity check on those actively odd-parity pairing (AOPP) two-bit pairs~\cite{xu2020sending,jiang2020zigzag,jiang2021composable}, then we can compress the bit-flip error significantly by post-selection, even if we use a large sending probability. Theories for finite-key effects have been presented~\cite{jiang2020zigzag,jiang2021composable} for the AOPP method and they make it clear that the AOPP method can make a high key rate and long distance even with the finite-key effects.
  So far the SNS protocol with AOPP method has been implemented in several experiments ~\cite{chen2020sending,pittaluga2021600,liu2021field,chen2021twin,chen2022quantum}, including the QKD experiments with the longest optical distance in the field test~\cite{chen2021twin}.

Here we present another method, the SNS protocol with redundant space (RS) method. By adopting the redundant space such as photon polarization in SNS protocol,  this method can also compress the bit-flip error rate significantly and it has a better performance than all prior art methods with a small data size. This makes it a good candidate protocol for the real application of the QKD task when fresh keys are demanded.

This paper is arranged as follows. In Sec.~\ref{protocol}, we introduce our RS method in detail, and the security proof of this new method can be found in Sec.~\ref{security proof}. Then we give numerical simulation results in Sec.~\ref{simulation}, and an improved protocol of this method is given in Sec.~\ref{improved}. This article ended with concluding remarks.

\section{SNS protocol with Redundant Space}~\label{protocol}
\begin{figure}[htbp]
	\centering
	\includegraphics[width=255pt]{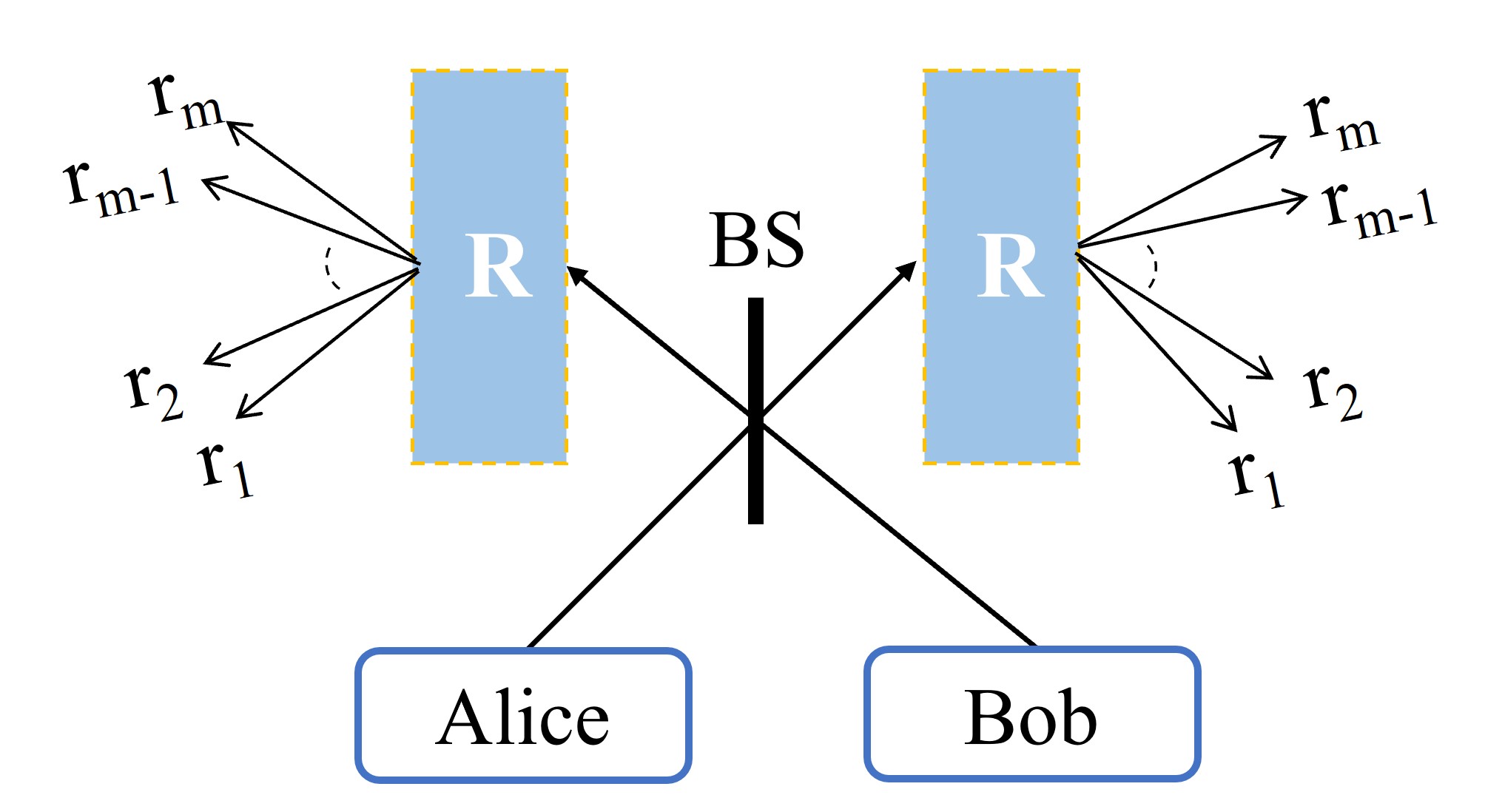}
	\caption{
		Schematic setup for SNS protocol with redundant space. 
		There are two groups of detectors, left and right after the beam splitter. Charlie is supposed to announce the heralded time windows with the information of which detector has clicked. This has actually announced both the side (left or right to the beam splitter) of the clicking detector and the specific $\bold R$ value ($\bold r_1$, $\bold r_2$ $…$ or $\bold r_m$) he has observed.
	}\label{fig:1e0}
\end{figure}

 Besides the Fock space used for the SNS protocol, we consider an RS method represented by physical quantity $\bold R$ which can be, e.g., polarization, spatial angular momentum, wavelength, time bin, and so on. 
 In each time window of the SNS protocol, Alice (Bob) randomly selects an $\bold R$ value from the set $\{\bold r_1, \bold r_2,\cdots \bold r_m\} (m\geq 2)$, if she (he) decides to send out a non-vacuum pulse. (If she (he) decides to send out vacuum at certain time window, she (he) just sends out nothing and there is no need to select an $\bold R$ value when sending out nothing.)
 By post-selection with specific $\bold R$, the bit-flip error is reduced and the performance of the protocol is improved. Say, at a certain time window, if Charlie’s measurement station is heralded with value $\bold r_1$ of $\bold R$ while Alice (Bob) has sent a non-vacuum coherent state with a different $\bold R$ value, e.g., $\bold r_2$, she (he) will request to discard the event of this time window. Surely, this type of post-selection with the RS method can reduce wrong bits.
   As given in Fig.~\ref{fig:1e0}, $\bold r_i$ indicates different $\bold R$ values that Charlie has observed.

\begin{figure}[htbp]
	\centering
	\includegraphics[width=195pt]{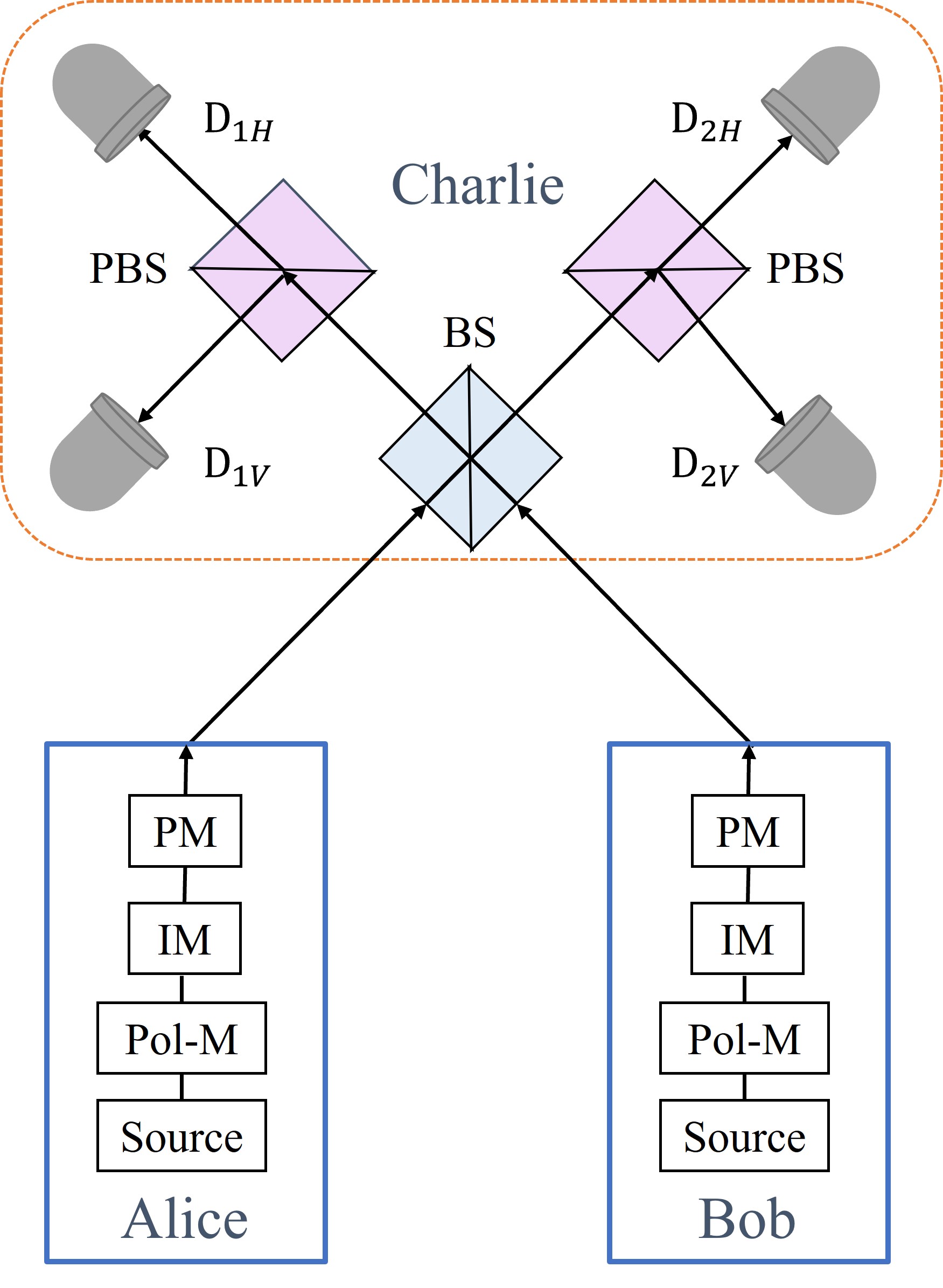}
	\caption{A schematic of the setup for SNS TF-QKD with Polarization mode protocol. PM: phase modulator; IM: intensity modulator; Pol-M: polarization modulator; BS: beam splitter; PBS: polarization beam splitter. $D_{1H}, D_{1V}, D_{2H}, D_{2V}$: single-photon detectors.  
		Charlie is supposed to announce those heralded time windows with the information of which detector has clicked. This has actually announced both the side (left or right to the beam splitter) of the clicking detector and the specific polarization ($H$ or $V$) he has observed.
	}\label{fig:1e2}
\end{figure}

For a better understanding of the idea of using the RS method, here we restate our main idea using the specific $\bold R$ of polarization mode. 
At any specific time window, if Alice  (Bob) decides to send a non-vacuum coherent state, she (he) will randomly choose its polarization from $H$ (horizontal polarization) or $V$ (vertical polarization). 
Charlie has two measurement ports after the beamsplitter, as shown in Fig.~\ref{fig:1e2}. Each of them measures the photon polarization, $H$ or $V$.
Charlie announces the one-detector heralded event and the polarization he has detected, i.e., Charlie announces the information of a specific detector at a specific port that has clicked. In the case that Charlie’s announced polarization of his measurement outcome is different from the polarization of the non-vacuum coherent state sent out by Alice (Bob), she (he) requests to discard that event.

We define a $lr$ time window: a time window when Alice sends out a pulse of intensity $\mu_l$ and Bob sends out a pulse of intensity $\mu_r$, 
where $l$,$r$ can be $x$, $y$, $z$, $v$. Here $v$ stands for a vacuum pulse with intensity $\mu_v=0$ for a time window.   Below is the complete protocol with the RS method of polarization mode:

{\bf Protocol 1}

{\em Step  1}. In each time window, Alice (Bob) randomly decides with probabilities $p_v$, $p_x$, $p_y$, $p_z$ to send out a vacuum pulse of intensity $\mu_v=0$, a non-vacuum pulse of intensity $\mu_x>0$, a non-vacuum pulse of intensity $\mu_y>\mu_x$, or a non-vacuum pulse of intensity $\mu_z>0$. Each of them also chooses the polarization $H$ or $V$ randomly with a probability of 50\% when she or he decides to send out a non-vacuum pulse (intensity $\mu_x$, $\mu_y$, or $\mu_z$). 

{\em Step 2}. Charlie announces those time windows with one-detector heralded events (we called them {\em heralded time windows}), and the specific information of which detector has clicked. Note that in announcing which detector has clicked, Charlie has actually announced both the side (left or right to the beam splitter) of the clicking detector and the polarization ($H$ or $V$) he has observed. 

{\em Step 3}. At any heralded time window, if Alice (Bob) has decided to send out a non-vacuum pulse and the polarization she (he) has chosen for the pulse is different from the measurement outcome of polarization announced by Charlie, she (he) requests to discard the event of this time window. The remaining heralded time windows  are defined as {\em accepted time windows}. And the corresponding events are defined as {\em accepted events}.

{\em Step 4}. \emph{They} (Alice and Bob) each announce those heralded time windows in which she (he) has chosen intensities $\mu_x$ and $\mu_y$. 
\emph{They} use the remaining survived events from those accepted time windows for code-bits. We can see that code-bits can only be generated from $vv,vz, zv,$ and $zz$ time windows.
Alice assigns a bit value of ``1'' if she has decided to send out a pulse of intensity $\mu_z$, and ``0'' if she has decided to send out a vacuum, i.e., intensity $\mu_v=0$. Bob's definition of bit value is opposite to that of Alice.
Through error tests, \emph{they} obtain the quantum bit error rate (QBER) value $E_t$ for code-bits.
With decoy-state analysis after error correction~\cite{hu2022universal}, \emph{they} can obtain the secure final key with the following key length formula~\cite{jiang2019unconditional}:  
\begin{equation}~\label{eq:keyrate2}
\begin{split}
N_f=&{n_1}[1-H( e^{ph}_1)]-fn_tH(E_t)\\
-&\log_2\frac{2}{\varepsilon_{cor}}-2\log_2\frac{1}{\sqrt{2}\varepsilon_{PA}\hat{\varepsilon}},
\end{split}
\end{equation}
where $H(x)=-x\log_2x-(1-x)\log_2(1-x)$ is the binary Shannon entropy function,  $f$ is the error correction efficiency factor, $ e_1^{ph}$ is the phase-flip error rate of untagged bits, and ${n_1}$ is the number of untagged bits. Values of ${n_1}$ and $ e_1^{ph}$ can be verified by decoy-state analysis, as shown in refs.~\cite{wang2018twin,yu2019sending,jiang2019unconditional}. 
Here $\varepsilon_{cor}$ is the failure probability of error correction, $\varepsilon_{PA}$ is the failure probability of privacy amplification, $\hat{\varepsilon}$ is the coefficient when using the chain rules for smooth min- and max-entropies~\cite{vitanov2013chain}. 
And $n_t$ is the number of code-bits, and value $E_t$ is the bit-flip error rate of those $n_t$ code-bits.
 In our protocol here, an event can be counted for {\em code-bit} if the following conditions are satisfied: 
 \begin{enumerate}
\item It is from an accepted time window;
\item Alice and Bob have chosen intensity $\mu_z$ or $\mu_v$;
\item Each one’s choice of intensity for the sent out pulse is never announced.
 \end{enumerate}
 {\bf Untagged bit}:  A code-bit is generated within a $zv$ or $vz$ time window by Alice or Bob sending a single-photon pulse while the other sends a vacuum pulse.
 
 \emph{They} need to announce the intensities of pulses \emph{they} each have used in those time windows which are not announced to be accepted time windows by Charlie. 
 In estimating $ e_1^{ph}$, \emph{they} have to set the phase slice condition~\cite{hu2019sending} in accepted events when both of them have chosen intensity $\mu_x$.  Note that $n_t$ and $E_t$ can be observed directly in the experiment.

\begin{remark}
	In Step 3 above, both Alice and Bob can decide to discard events when their polarization of pulse of intensity $\mu_z$ is different from Charlie’s observed polarization. This post-selection plays a key role in compressing the bit-flip error rate of those $n_t$ survived code-bits.  Those wrong bits from heralded $zz$ time windows are for sure rejected by this step if Alice and Bob have chosen different polarization. Here the result of bit-flip error rate compressing alone is not as effective as that of the AOPP method~\cite{xu2020sending}, but the phase-flip error rate does not rise as what happens in the AOPP method. 
	Moreover, the RS method proposed here does not consume too many raw bits in compressing QBER, as what happens in the AOPP method.
	This makes it possible for our protocol to produce advantageous results under certain conditions, especially when the fresh key can only be distilled from a small data size.
\end{remark}

Here, we give intuitive explanations of how this protocol with redundant polarization space reduces the QBER value. In the original SNS protocol~\cite{wang2018twin}, if we ignore the vacuum counts, the observed QBER value is:
\begin{equation}\label{eq:2}
E_{t0}\approx\frac{p_z^2S_{zz}}{p_z^2S_{zz}+p_zp_v(S_{vz}+S_{zv})},
\end{equation}
where $S_{\alpha\beta}$ is the counting rate for $\alpha\beta$ time windows, and $\alpha, \beta \in \{z,v\}$. As defined in ref.~\cite{wang2018twin}, $n_{\alpha\beta}$ indicates the total number of one-detector heralded events in $\alpha\beta$ time windows, and $N_{\alpha\beta}$ indicates the total number of $\alpha\beta$ time windows, then $S_{\alpha\beta}=n_{\alpha\beta}/N_{\alpha\beta}$. 
In protocol 1 above, the photon polarization modes $H$ and $V$ make a two-dimensional redundant space. Since the heralded time windows with different polarization from two sides (Alice and Bob) are discarded in the protocol, normally, about half of heralded $zz$ time windows would be discarded.
Therefore, if we ignore the vacuum counts, the observed QBER among all code-bits would be
\begin{equation}\label{eq:3}
E_t=E_{t}(2)\approx \frac{p_z^2S_{zz}/2}{p_z^2S_{zz}/2+p_zp_v(S_{vz}+S_{zv})}.
\end{equation}
This means $E_{t}(2)$ is compressed to a value of only a bit larger than half of $E_{t0}$. Here ``2''  in the parenthesis indicates 2 modes of redundant polarization space. 
If we use redundant space with $m$ modes, the observed QBER value is:
\begin{equation}\label{QBER-n}
E_t=E_{t}(m)\approx \frac{p_z^2S_{zz}/m}{p_z^2S_{zz}/m+p_zp_v(S_{vz}+S_{zv})}.
\end{equation}
This further compresses the QBER value to $1/m$ of that without using our RS method, represented by Eq.~\eqref{eq:2}.
A more exact formula for QBER value with the RS  method is shown by Eq.~\eqref{eq:simulate} and Appendix~\ref{appendixB}, which takes the vacuum counts into account.

\section{security proof}~\label{security proof}

Here we shall show the MDI security for our protocol with the RS method as presented in Sec.~\ref{protocol}.
The original SNS protocol~\cite{wang2018twin} takes MDI security and it is not limited to any specific additional space.  Without loss of generality, we consider the security of this new protocol with polarization mode as an example. 

{We shall start with virtual protocols}.
Protocol $\mathcal{H}$, a bit-flip error-free protocol using polarization ${H}$ for non-vacuum pulses. With secret discussion in advance, there are only time windows when both of them send out pulses of intensity $\mu_x$ and time windows when one side sends out pulses of intensity $\mu_z$ and the other side sends out a vacuum. In a 4-intensity protocol, there are also time windows one side sends out pulses of intensity $\mu_y$, and the other side sends out a vacuum. At all time windows, the non-vacuum pulses are prepared in polarization $H$.  In protocol $\mathcal{H}$, \emph{they} take post-selection using polarization information. In particular, after sending out quantum states, there are the following steps for classical communication and post-selection:

{\em Step 1.} Charlie announces those heralded time windows and the polarization ($H$ or $V$) he has observed, and also the position (right or left to the beam splitter) of the heralded detector.

{\em Step 2.} Given any heralded time window with a polarization $V$ announced by Charlie, any party (Alice or Bob) who has sent out the non-vacuum pulse in the time window will announce ``No''; given any heralded time window with a polarization $H$ announced by Charlie, \emph{they}  keep silent.

{\em Step 3.} \emph{They} take post-selection of events which are from heralded time windows announced by Charlie in Step 1 and neither Alice nor Bob announces ``No'' in Step 2 (i.e., both of them keep silent in Step 2).  \emph{They} take post-processing on the post-selected events and distill the final key with the following key length formula
\begin{equation}\label{eqkey:H}
\begin{split}
n_{\mathcal{H}} = {n_1}(\mathcal{H})\left[1-H( e^{ph}_1(\mathcal{H}))\right].
\end{split}
\end{equation}
Here $n_1(\mathcal H)$ is the number of untagged bits from protocol $\mathcal H$, and $e^{ph}_1(\mathcal{H})$ is their phase-flip error rate.

We can easily see the security of this protocol by comparing it with the original SNS protocol $\mathcal{O}$ where Charlie alone takes the post-selection by announcing the correctly heralded time windows after quantum communication. In general, protocol $\mathcal{O}$ is secure with whatever Charlie, say, Charlie can actually choose whatever subset of time windows for his announcement of ``correctly heralded time windows''.
 The security of protocol $\mathcal{H}$ is obviously equivalent to $\mathcal{O}$ with a specific Charlie who announces the same post-selection taken by Alice and Bob in protocol $\mathcal{H}$ as the correctly heralded time windows. 
Note that the secure key rate is only dependent on parameters such as $e_1^{ph}$ of those post-selected code-bits and hence we don't need to worry about the classical communications taken in Step 2 of protocol $\mathcal{H}$ because it only announces the information of the discarded bits. Neither do we need to worry about the announced information in Step 1 of protocol $\mathcal{H}$, since the original SNS protocol allows whatever actions of Charlie.

Replacing all polarization $H$ by polarization $V$ in protocol $\mathcal{H}$ above, we have protocol $\mathcal{V}$, which is the bit-flip error-free protocol with polarization $V$. It has the following key length formula:
 \begin{equation}
 \begin{split}
 n_{\mathcal{V}} = {n_1}(\mathcal{V})\left[1-H( e^{ph}_1(\mathcal{V}))\right],
 \end{split}
 \end{equation}
 where $n_1(\mathcal V)$ is the number of untagged bits from protocol $\mathcal V$, and $e^{ph}_1(\mathcal{V})$ is their phase-flip error rate. Similarly, protocol $\mathcal{V}$ is also MDI secure.
 
 Also, mixing the two protocols above, we have protocol $\mathcal{M}$, where in some time windows \emph{they} use protocol $\mathcal{H}$, and in other time windows, \emph{they} use protocol $\mathcal{V}$. 
 (In this virtual protocol $\mathcal{M}$, \emph{they} can take secret discussion in advance on the protocol choice and state choice for each time window.) 
 \emph{They} use Step 2 and Step 3 of protocol 1 in Sec.~\ref{protocol} to do post-selection. In particular, Charlie announces heralded time windows with polarization $H$ or $V$ he has observed. Given any heralded time window announced by Charlie, any party (Alice or Bob) who has sent out a non-vacuum pulse in the time window announces rejection if the polarization of the sent out non-vacuum pulse is different from the polarization announced by Charlie, otherwise, he or she keeps silent. If both Alice and Bob keep silent, the event of the corresponding heralded time window is accepted. The secure key length formula is 
 \begin{equation}
 \begin{split}
 N_f = n_{\mathcal{H}}+n_{\mathcal{V}}.
 \end{split}
 \end{equation}
 Of course, \emph{they} can replace the phase-flip error rate of each protocol with the averaged phase-flip error rate and use the following key length formula: 
 \begin{equation}\label{key-M}
 \begin{split}
 N'_f=\tilde {n}_1 \left[1- H(\tilde e_1^{ph})\right],
 \end{split}
 \end{equation}
 where
 \begin{equation}
 \begin{split}
 \tilde n_1=&{n_1}(\mathcal{H})+{n_1}(\mathcal{V}),\\
\tilde  e_1^{ph}=&\frac{e_1^{ph}(\mathcal{H}){n_1}(\mathcal{H})+e^{ph}_1(\mathcal{V}){n_1}(\mathcal{V})}{\tilde n_1}.
 \end{split}
 \end{equation}
  We can use this because inequality $N_f \ge N'_f$ always holds mathematically due to the concavity of the entropy function.
 Using this formula, \emph{they} do not have to know which bit belongs to which protocol. 
 
 {Now consider the real protocol} where each side takes probabilities to send different types of intensities with a random choice of polarization $H$ or $V$ respectively. 
 In the protocol, after Charlie’s announcement, \emph{they} take post-selection by Step 3 of protocol 1 in Sec.~\ref{protocol}, then \emph{they} each announced those time windows she or he has sent out pulses of intensity $\mu_x$ or $\mu_y$ and the corresponding polarization. 
  After post-selection in the real protocol, the survived events from time windows of $vv$, $zz$, $vz$, $zv$ after post-selection include 2 kinds of subsets:
 subset $\mathcal{C}_H \oplus \mathcal{C}_V$, which is the post-selected data from the mixed protocol $\mathcal M$, and subset $\mathcal T$ includes all other data. 
 Subset $\mathcal{C}_H \oplus \mathcal{C}_V$ contains accepted events from $zv$ time windows and $vz$ time windows only, which has no bit-flip error. 
 In particular, $\mathcal T$ includes accepted events from $vv$ time windows and accepted events from $zz$ time windows with the same polarization,
  which shall be treated as tagged bits. Note that not all one-detector heralded events from $zz$ time windows can survive from the post-selection. Those one-detector heralded events from $zz$ time windows when Alice and Bob have chosen different polarization have been rejected in the post-selection. 
  
  Although \emph{they} need secret discussion in virtual protocol $\mathcal{H}$ or $\mathcal{V}$, \emph{they} don't need to do so in the real protocol. Suppose in our real protocol, each side takes probabilities $p_0$, $p_{Hx}$, $p_{Hy}$, $p_{Hz}$, $p_{Vx}$, $p_{Vy}$, $p_{Vz}$, to send out a vacuum, $\mu_x$ with polarization $H$, $\mu_y$ with polarization $H$, $\mu_z$ with polarization $H$, $\mu_x$ with polarization $V$, $\mu_y$ with polarization $V$, $\mu_z$ with polarization $V$, respectively and independently. This means any time window has a probability $P_H = 2 (p_0 p_{Hy} +p_0p_{Hz}) + p_{Hx}^2$ belongs to protocol $\mathcal{H}$, and probability $P_V= 2 (p_0 p_{Vy} +p_0p_{Vz})+ p_{Vx}^2$  belongs to protocol $\mathcal{V}$.
  Say, there are subsets $\mathcal{C}_H$ and $\mathcal{C}_V$ of time windows which send out states for protocol $\mathcal{H}$ and protocol $\mathcal{V}$, respectively.
  The result would be secure if \emph{they} only use the post-selected data from $\mathcal{C}_H \oplus \mathcal{C}_V$ to calculate the final key rate. Although \emph{they} don't know which data are from $\mathcal{C}_H \oplus \mathcal{C}_V$, \emph{they} can use the tagged model with the decoy-state method to make their result as secure as the case \emph{they} have only used data from $\mathcal{C}_H \oplus \mathcal{C}_V$.
  
  If \emph{they} only use data of subset $\mathcal{C}_H \oplus \mathcal{C}_V$, \emph{they} can simply apply Eq.~\eqref{key-M} for the final key length, which is secure.
 Regarding data in subset $\mathcal T$ as tagged bits, \emph{they} can distill the secure final keys with the key length formula:
 \begin{equation}\label{key-secure}
 \tilde N_f=\tilde  n_1\left[1-H(\tilde e_1^{ph})\right] - f \tilde n_t H(\tilde E_t).
 \end{equation}
 The notation $\tilde n_t$ is for the total number of bits in subsets $\mathcal{C}_H \oplus \mathcal{C}_V$ and $\mathcal T$, and the notation $\tilde E_t$ is for the bit-flip error rate of these $\tilde n_t$ bits.  Here we can see that $\tilde{n}_1 $ is the total number of untagged bits, and if the finite-key effect is considered, the key length formula is just Eq.~\eqref{eq:keyrate2}.
 
  Thus the security of this new protocol with polarization mode has been proved. Obviously, this security proof process can be applied to the 
 more general redundant space cases using whatever physical quantity $\bold R$, which completes our security proof.

\section{numerical simulation}~\label{simulation}


In the following, we use the linear model~\cite{jiang2019unconditional} with standard optical fiber (0.2 dB/km) to simulate the observed values in experiments.
 Without loss of generality, we consider a symmetrical channel,  also the intensity of pulses and the corresponding channel loss are the same in different additional space.
 In previous works, the original SNS protocol~\cite{wang2018twin} has been improved, especially by the AOPP method~\cite{xu2020sending,jiang2020zigzag,jiang2021composable}, which has been adopted in several experiments~\cite{chen2020sending,pittaluga2021600} through laboratory optical fiber and created the record for field tests~\cite{liu2021field,chen2021twin} of all types of fiber-based QKD systems. We compare this work with AOPP, and the finite-key effects have been taken into consideration. In this work, We take $\xi=10^{-10}$ as the failure probability of Chernoff Bound ~\cite{chernoff1952measure,curty2014finite}, 
 and we set $\varepsilon_{cor}=\hat{\varepsilon}=\varepsilon_{PA}=\xi$. 
 The key rate formula is given in Eq.~\eqref{eq:keyrate2}, where values of ${n_1}$ and $ e_1^{ph}$ can be verified through decoy-state analysis, as shown in Appendix~\ref{appendixA}, and both the QBER value $E_t$ and the number of code-bits $n_t$ would be directly observed (tested) in experiments. Mathematically, the $n_t$ and $E_t$ for RS method with $m$ modes are:
  \begin{equation}\label{eq:simulate}
  \begin{split}
   n_t=&n_V+n_C+n_D,\\
  E_t=&\frac{n_V+n_D}{n_V+n_C+n_D}.
   \end{split}
  \end{equation}
 Here $n_V$ is the total number of accepted events in $vv$ time windows, $n_D$ is the total number of accepted events in $zz$ time windows, and $n_C$ is the total number of accepted events in both $vz$ time windows and $zv$ time windows.
 Here in our numerical simulation, we use Eq.~\eqref{eq:simulate} to estimate the QBER value which would be observed (tested) directly in an experiment.
 Details of the calculations of QBER value are given in Appendix~\ref{appendixB}.

 \begin{table}[htb]
 	\begin{ruledtabular}
 		\begin{tabular}{lllll}
 			& $d$          & $\eta_0$     &$E_d$       &$N$   \\ \hline
 			A &$1\times10^{-8}$      &$0.50$      &$0.03$     &$1\times 10^{8}$\\ 
 			B &$1\times10^{-8}$      &$0.50$     &$0.03$      &--- \\
 			C &$1\times10^{-9}$      &$0.50$     &$0.03$      &$1\times 10^{10}$ \\
 			D &$1\times10^{-8}$      &$0.50$      &$0.03$     &$1\times 10^{12}$\\ 
 		\end{tabular}
 	\end{ruledtabular}
 	\caption{Device parameters used in numerical simulations.
 		$d$: the dark count rate.
 		$\eta_0$: the detection efficiency of all detectors.
 		$E_d$: the misalignment error.
 		$N$: total number of time windows.
 		The fiber loss is set as $0.2$ dB/km, and we set the error correction efficiency factor as $f=1.1$ in this work.
 	}\label{tab:parameters}
 \end{table}

\begin{figure}[htbp]
	\includegraphics[width=285pt]{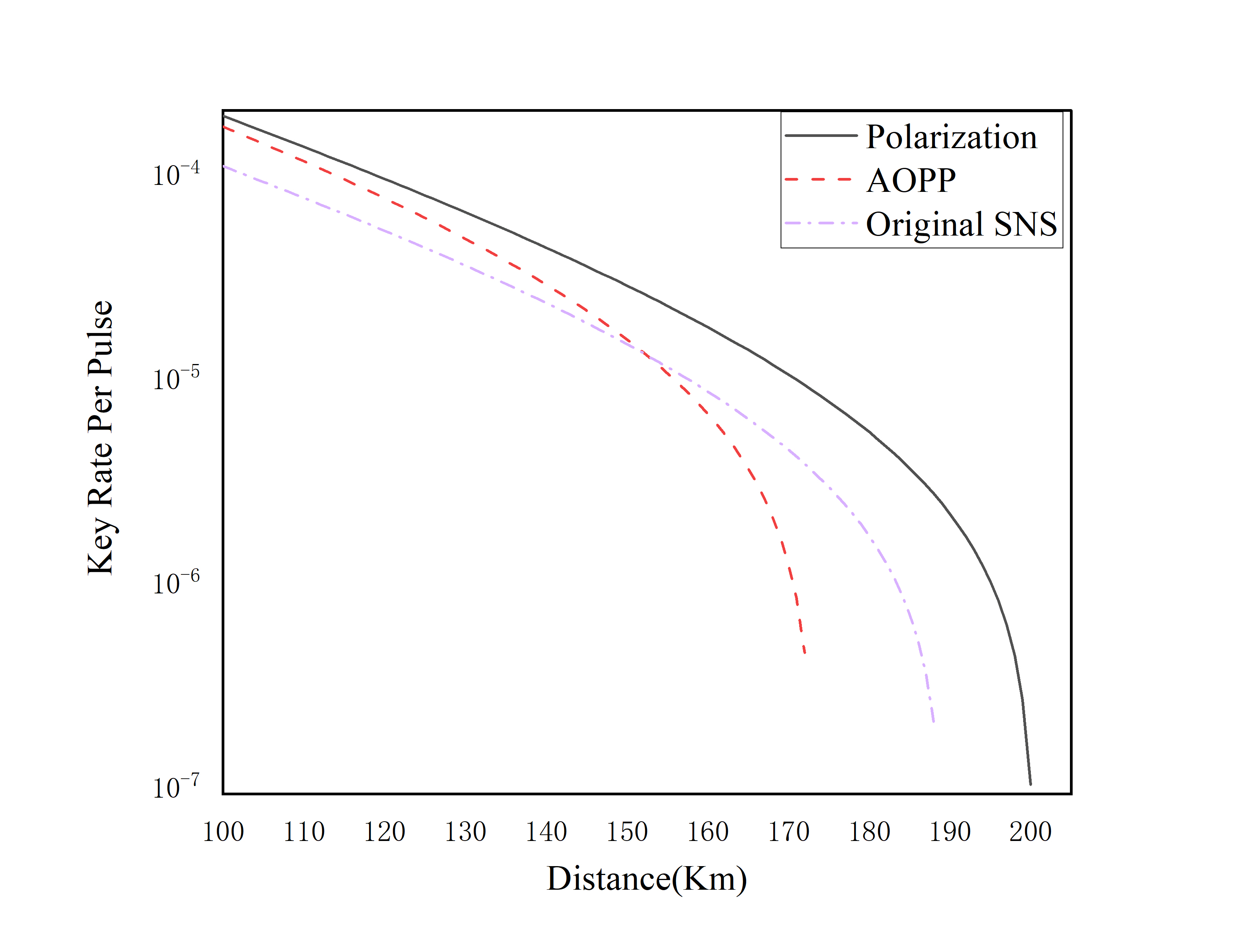}
	\caption{(Color online)  The optimized  key rates of RS method of polarization by Eq.~\eqref{eq:keyrate2} (dark solid line), the original SNS protocol ~\cite{wang2018twin} (purple dot-dash line), and the AOPP method~\cite{xu2020sending,jiang2021higher} (red dash line). 
		Devices' parameters are given by row A of Table ~\ref{tab:parameters}. }\label{fig:1e3}
\end{figure}

In Fig.~\ref{fig:1e3} we compare the secure key rate of the RS method of polarization given by Eq.~\eqref{eq:keyrate2} (the dark solid line), the original SNS protocol ~\cite{wang2018twin} (purple dot-dash line), and the AOPP method~\cite{xu2020sending,jiang2021higher} (red dash line). The simulation results show that, given a small data size ($N\sim 10^{8}$), the secure key rate of the new protocol we present here, is about 134\%  higher than that of the original SNS protocol~\cite{wang2018twin} at the point of 170km, and the secure distance is nearly 30 km longer compared with AOPP method~\cite{xu2020sending,jiang2021higher}. Parameters are given in line A of Tab.~\ref{tab:parameters}.


\begin{figure}[htbp]
	\includegraphics[width=285pt]{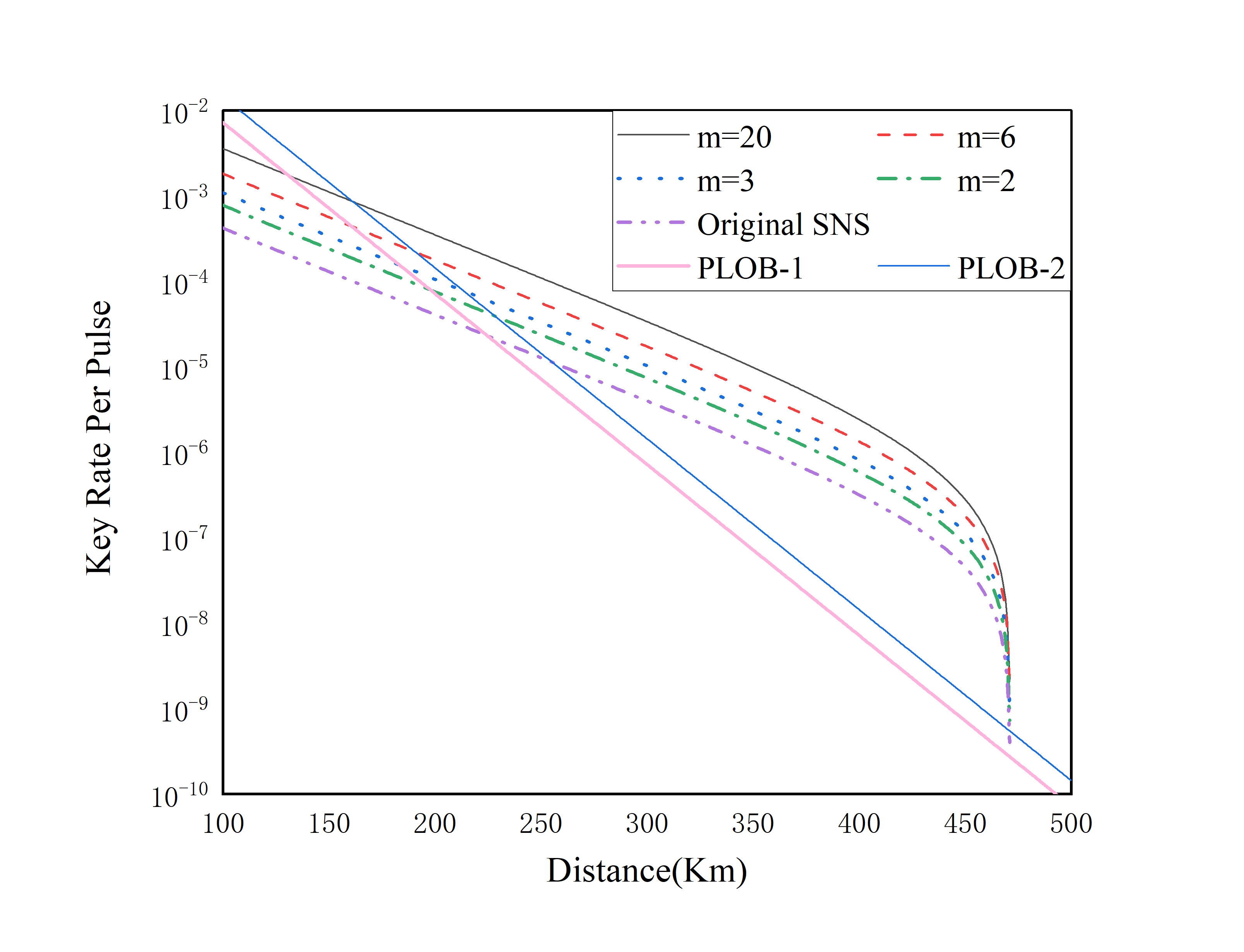}
	\caption{(Color online)  Asymptotic  key rates of RS method of WDM,  the total number of wavelength divisions is $m=20$ (dark solid line), $m=6$ (red dash line), $m=3$ (blue dotted line) and $m=2$ (green dot-dash line) respectively, and that given by the original SNS protocol ~\cite{wang2018twin} (dash-dot-dot line).  PLOB-1 is repeater-less key rate bound ~\cite{pirandola2017fundamental} with detector efficiency $\eta_0=0.5$, i.e., the relative limit of repeater-less key rate. PLOB-2 is repeater-less key rate bound  with detector efficiency $\eta_0=1$, i.e., the absolute limit of repeater-less key rate.
	Devices' parameters are given by row B of Table ~\ref{tab:parameters}. }\label{fig:1e4}
\end{figure}

In Fig.~\ref{fig:1e4} we compare the secure key rate of the RS method of wavelength division in an asymptotic case (the total number of wavelength divisions is $m=20$, $m=6$, $m=3$, and $m=2$ respectively), with that given by the original SNS protocol ~\cite{wang2018twin} (dash-dot-dot line). Note that the result of $m=2$ is also  the result of RS  with photon polarization. The simulation results show that the secure key rate of the new protocol increases obviously with the increase of the total number of different wavelength divisions. PLOB-1 is the repeater-less key rate bound ~\cite{pirandola2017fundamental} with detector efficiency $\eta_0=0.5$, i.e., the relative limit of the repeater-less key rate. PLOB-2 is a repeater-less key rate bound  with detector efficiency $\eta_0=1$, i.e., the absolute limit of the repeater-less key rate.
The absolute PLOB bound and the relative PLOB bound are the bounds with whatever devices and the practical bound assuming
the limited detection efficiency, respectively. At the distance of  300 km, the secure key rate of $m=2$ is about $80\%$ higher than the secure key rate of the original SNS protocol~\cite{wang2018twin}, and this number for $m=3$, $m=6$, and $m=20$ are about $150\%$, $300\%$, and $700\%$ respectively. Parameters are given in line B of Tab.~\ref{tab:parameters}.

\begin{figure}[htbp]
	\includegraphics[width=285pt]{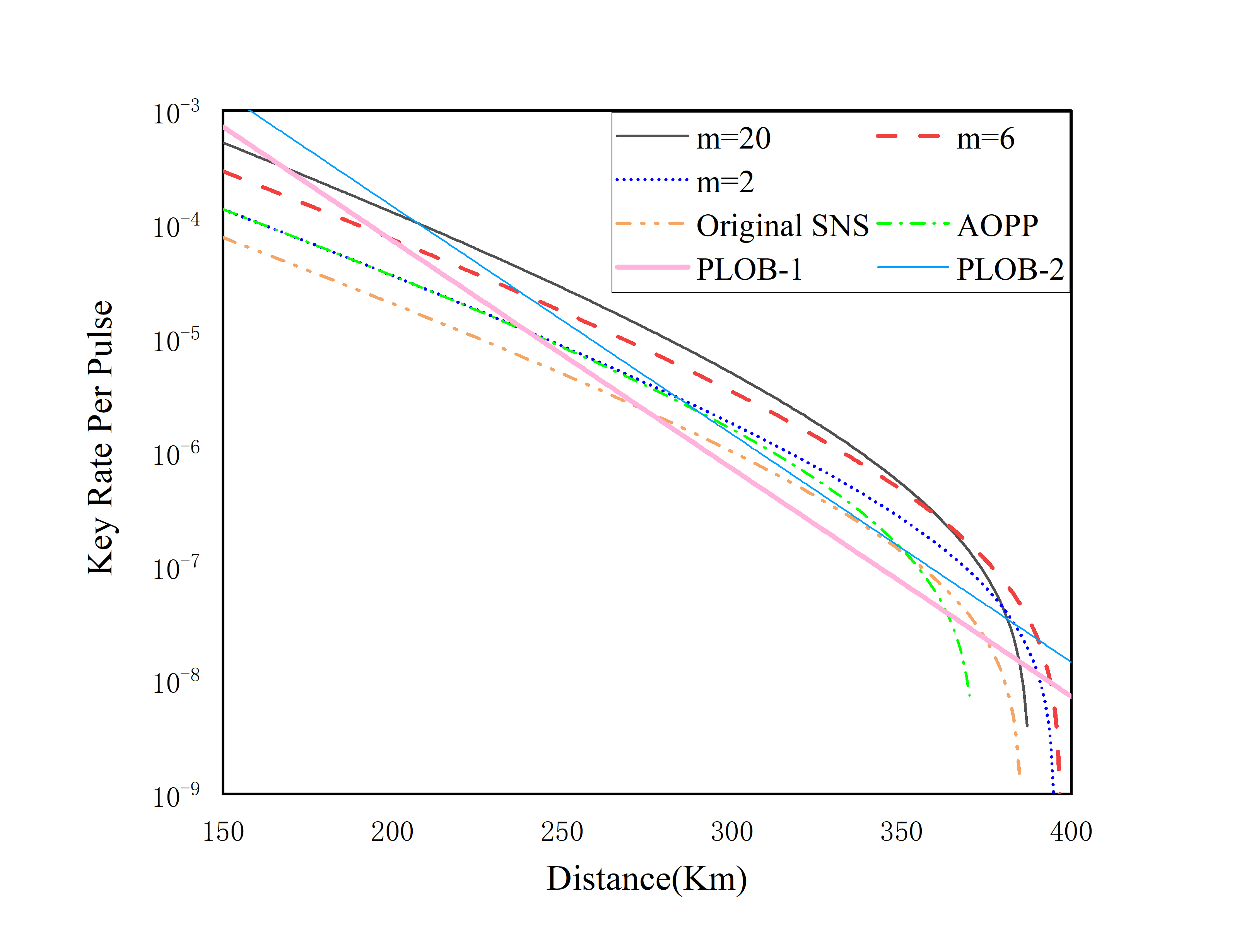}
	\caption{(Color online)  
	The optimized  key rates of RS method of WDM given by Eq.~\eqref{eq:keyrate2},  the total number of wavelength divisions is $m=20$ (dark solid line), $m=6$ (red dash line),
	and $m=2$ (blue short dotted line) respectively, and the original SNS protocol ~\cite{wang2018twin} (brown dash-dot-dot line), and AOPP method~\cite{jiang2021composable} (green dot-dash line). PLOB-1 is repeater-less key rate bound ~\cite{pirandola2017fundamental} with detector efficiency $\eta_0=0.5$, i.e., the relative limit of the repeater-less key rate. PLOB-2 is a repeater-less key rate bound  with detector efficiency $\eta_0=1$, i.e., the absolute limit of the repeater-less key rate.
		Devices' parameters are given by row C of Table ~\ref{tab:parameters}. }\label{fig:1e7}
\end{figure}

\begin{figure}[htbp]
	\includegraphics[width=285pt]{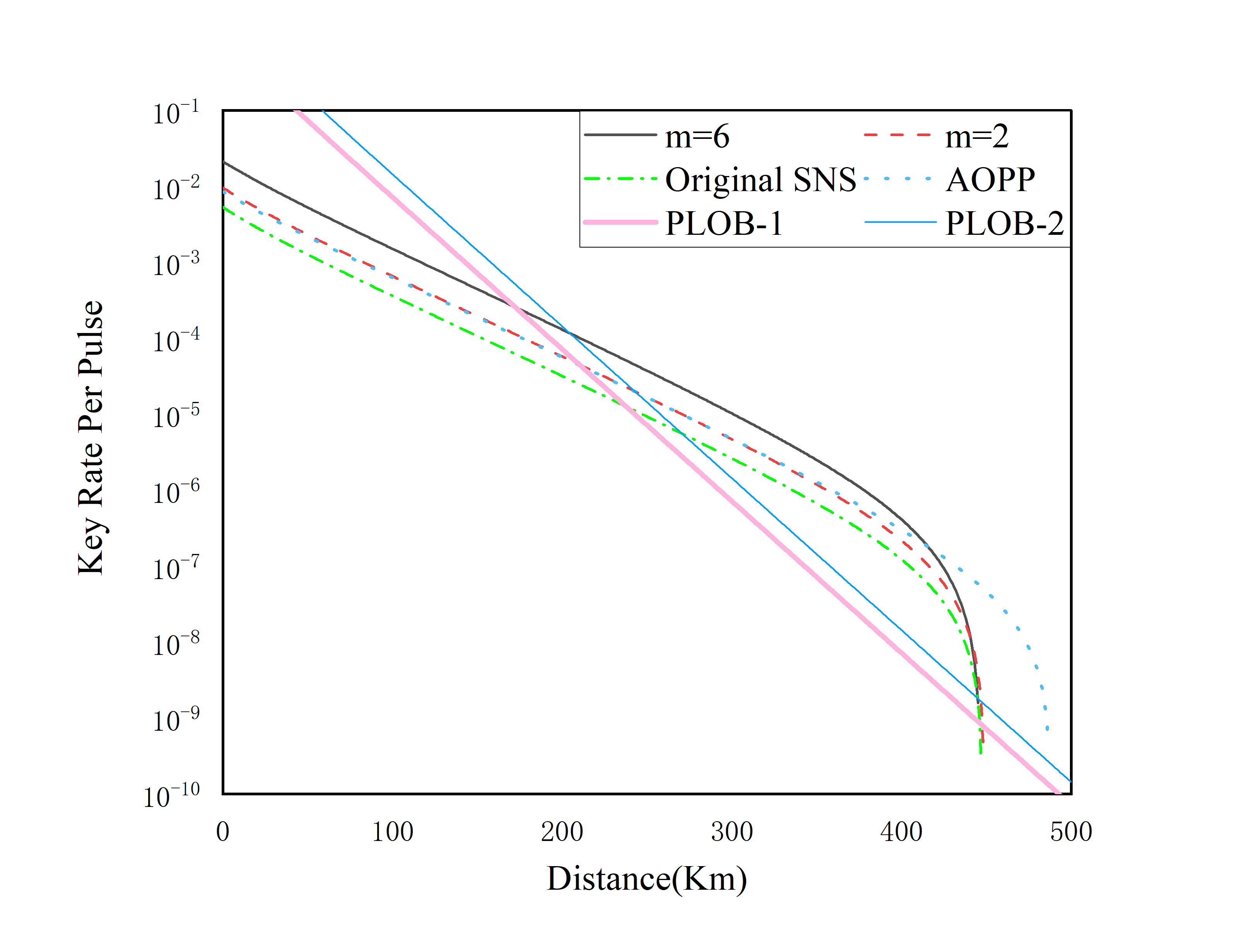}
	\caption{(Color online)  
		The optimized  key rates of the RS method of WDM given by Eq.~\eqref{eq:keyrate2},  the total number of wavelength divisions is $m=6$ (dark solid line),  $m=2$ (red dash line) respectively, and the original SNS protocol ~\cite{wang2018twin} (green dot-dash line), and the AOPP method~\cite{jiang2021composable} (blue dotted line). PLOB-1 is a repeater-less key rate bound ~\cite{pirandola2017fundamental} with detector efficiency $\eta_0=0.5$, i.e., the relative limit of the repeater-less key rate. PLOB-2 is a repeater-less key rate bound  with detector efficiency $\eta_0=1$, i.e., the absolute limit of the repeater-less key rate.
		Devices' parameters are given by row D of Table ~\ref{tab:parameters}. }\label{fig:1e8}
\end{figure}

\begin{table}[htb]
	\begin{ruledtabular}
		\begin{tabular}{l|l|l|l}
			& 250 km  & 300 km  &350km\\
			\hline
			PLOB-2~\cite{pirandola2017fundamental}  &$1.44\times 10^{-5}$  &$1.44\times 10^{-6}$  &$1.44\times 10^{-7}$\\
			SNS~\cite{wang2018twin} &$4.90\times 10^{-6}$   &$1.01\times 10^{-6}$  &$1.34\times 10^{-7}$         \\      
			AOPP~\cite{jiang2021composable}&$8.38\times 10^{-6}$  &$1.59\times 10^{-6}$  &$1.43\times 10^{-7}$\\ 
			$m=2$  &$8.57\times 10^{-6}$  &$1.79\times 10^{-6}$ &$2.62\times 10^{-7}$\\
			$m=6$ &$1.72\times 10^{-5}$   & $3.41\times 10^{-6}$  &$4.66\times 10^{-7}$ \\
			$m=20$ &$2.96\times 10^{-5}$  &$4.94\times 10^{-6}$  &$5.30\times 10^{-7}$\\
		\end{tabular}
	\end{ruledtabular}
	\caption{ Comparison of the key rates at three distance points given by different methods in Fig.~\ref{fig:1e7}. 
		Devices' parameters are given by row C of Table ~\ref{tab:parameters}.
	}\label{tab:parameters2}
\end{table}

In Fig.~\ref{fig:1e7} we compare the secure key rate of the RS method of WDM given by Eq.~\eqref{eq:keyrate2} (total number of wavelength divisions is $m=20$, $m=6$, and $m=2$ respectively), with that given by the original SNS protocol ~\cite{wang2018twin} and AOPP method.  We can see that the secure key rate of the new protocol increases with the increase of the total number of wavelength divisions. But when this number becomes larger than $m=20$, the secure distance becomes shorter, for the influence of dark count and the statistical flctuation in calculating  phase-flip error rate becomes obvious.

 As shown in Table~\ref{tab:parameters2}, we compare the key rates given by different methods at the distance of  250 km, 300 km, and 350 km. We can see that the secure key rate of the new protocol increases with the increase of the total number of modes in redundant space, and the key rate given by this method can largely exceed the absolute PLOB bound even giving a small number of pulses.  
At the distance of 350 km, the secure key rate of the RS method of WDM with $m=2$ is about $96\%$ higher than the secure key rate of the original SNS protocol~\cite{wang2018twin}, and this number for $m=6$, and $m=20$ are about, $248\%$ and $296\%$ respectively. Parameters are given in line C of Tab.~\ref{tab:parameters}.

In Fig.~\ref{fig:1e8} we compare the secure key rate of the RS method of wavelength division given by Eq.~\eqref{eq:keyrate2} (total number of wavelength divisions is $m=6$, $m=2$ respectively), with that given by the original SNS protocol ~\cite{wang2018twin} and the AOPP method~\cite{jiang2021composable}.  At a certain distance, when the total number of wavelength divisions is $m=2$, the secure key rate of the new protocol is quite close to the secure key rate of the AOPP method, and when the total number of wavelength divisions increases to $m=6$, the secure key rate of this new protocol is roughly twice that of the AOPP method. However, when the date size is not small ($10^{12}$), the secure distance of this new protocol becomes shorter than that of the AOPP method. Parameters are given in line D of Tab.~\ref{tab:parameters}.

The numerical simulation above shows clear performance for our protocol. With whatever channel disturbance, our result is still MDI secure, also Alice and Bob will test the Quantum bit error rate (QBER) after the post-selection of those time windows. 
We believe the performance advantage still exists if the channel disturbance is not too large.
The advantage comes from the fact that the redundant space used in our protocol can suppress QBER in code-bits significantly, and it would not raise the phase-error rate compared with previous art results of bit-flip error suppression such as the AOPP method.
Thus, the RS method has better performance under the small data size condition, and this also means that our protocol can work with fairly large channel noise. 
It is an interesting problem to study the performance of our protocol under more practical conditions with polarization drifts~\cite{ wang2015phase, wang2017measurement, lu2022unbalanced,Fan2022robust,laing2010reference}.

\begin{figure}[htbp]
	\includegraphics[width=285pt]{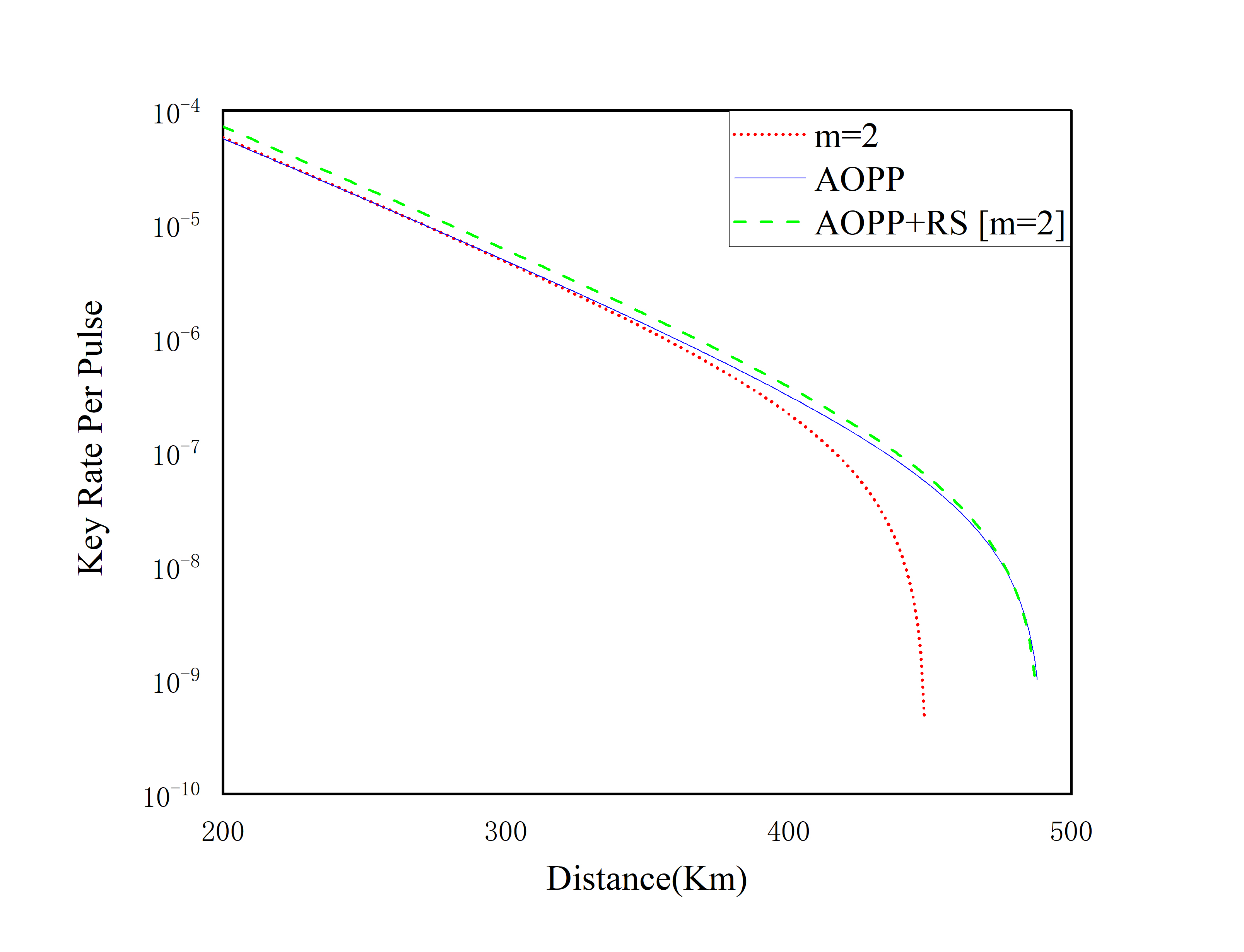}
	\caption{(Color online)  
		The optimized  key rates of the RS method of polarization given by Eq.~\eqref{eq:keyrate2} (red short-dotted line), and the AOPP method ~\cite{jiang2021composable} (blue solid line), and the AOPP method combined with RS method (green dash line).
		Devices' parameters are given by row D of Table ~\ref{tab:parameters}. }\label{fig:1e9}
\end{figure}

Naturally, one may try to combine the RS method here and the AOPP method (AOPP+RS) to obtain an even more advantageous result. Indeed, under certain specific conditions such as Fig.~\ref{fig:1e9},  the direct combination, AOPP+RS, can lead to advantageous results to RS or AOPP alone. But under some other conditions, AOPP+RS does not produce any advantageous results.
Some subtle design is necessary to better combine the advantages of both methods. This may enable us to propose a new protocol that is more efficient in both secure distance and key rates in general conditions.

\section{Improved Protocol}\label{improved}
 In our calculations above we have used our protocol in Sec.~\ref{protocol}, which takes the same decoy-state analysis method as the prior works~\cite{jiang2019unconditional,jiang2021composable,yu2019sending} for a fair comparison. Definitely, the recently proposed improved method of SNS protocol using decoy-state analysis after error correction~\cite{hu2022universal} can also apply to the RS method in this work. 
 
 Combined with the method proposed in ref.~\cite{hu2022universal}, Alice and Bob do not announce the intensity \emph{they} choose in heralded time windows, and all kinds of  accepted events can be used for code-bits. Alice (Bob) regards all accepted events when she (he) uses the vacuum as a bit value ``0'' (``1''), and assigns a bit value ``1'' (``0'') for  the rest (intensity $\mu_l$, $l\in \{x,y,z\}$). The source of the untagged bits is not limited to signal pulses (intensity $\mu_z$). It extends to all $l v$ time windows and ${v}r$ time windows, with $l,r\in \{x,y,z\}$.
  After the quantum communication part, Alice and Bob have $n'_t$ accepted time windows, and \emph{they} perform error correction first, then \emph{they} apply decoy-state analysis to verify the total number and the phase-flip error rate of untagged bit.
 
  In this case, the announcements in Step 4
  of our protocol in Sec.~\ref{protocol} is not necessary and the key rate can be further improved. 
  \emph{They} can choose all accepted events for code-bits to distill the final keys, thus the number of untagged bits is improved compared with the original SNS protocol, while the bit-flip error rate is enlarged. Thus, \emph{they} can also choose a part of accepted events for code-bits, i.e., limiting $l, r \in \{x,y\}$, $l, r \in \{y,z\} $ or $l,r\in \{z\}$ for code-bits, which is a trade-off.
  The key length formulas such as Eq.(2), Eq.(20), and Eq.(23) in ref.~\cite{hu2022universal} can all be used here.

\section{conclusion}
Based on the SNS TF-QKD protocol~\cite{wang2018twin}, we propose a new protocol with redundant space. Redundant space like polarization mode and wavelength division are given here as examples. Our simulation results show that our new method can achieve better performance at all distance points over the original SNS protocol. Further, when the data size is small like $10^{8}$, this RS method can improve the maximum distance of the AOPP method by nearly 30 km, and greatly improve the key rate at long distances. Our protocol can also apply to efficient quantum digital signature by taking the post-processing method such as refs.~\cite{amiri2016secure,qin2022quantum}. This will be reported elsewhere.

\section{acknowledgment}
We acknowledge the financial support in part by Ministration of Science and Technology of China through National Natural Science Foundation of China Grant Nos.11974204, 12174215, 12104184, 12147107; Open Research Fund Program of the State Key Laboratory of Low-Dimensional Quantum Physics Grant No. KF202110; Key R$\&$D Plan of Shandong Province Grant Nos. 2021ZDPT01.

 \begin{appendices}
 \section{Decoy-state analysis}\label{appendixA}
 In this section, we give the method of decoy-state analysis. 
For general redundant space adopting cases, suppose the physical quantity $\bold R$ has $m$ elements, that is $\{\bold r_1,\bold r_2,\cdots, \bold r_m\}, m\ge 2$.  
For ease of presentation, we make the following notations:\\
$P_{r_j}$: the probability when $\bold r_j$ is chosen by Alice or Bob;\\
$N_{lr}(r_j)$: the number of $lr$ time windows when both Alice and Bob choose $\bold r_j$ ($\mu_l\neq 0$ and $\mu_r\neq 0$) or one of them chooses $\bold r_j$ with a non-vacuum pulse, while the other sends a vacuum;\\
$N_{vv}$: the total number of time windows when both Alice and Bob choose to send out a vacuum;\\
$n_{lr}(r_j)$: the number of accepted events in $lr$ time windows when Charlie announces the value of $\bold r_j$;\\
$S_{lr}(r_j)$: the counting rate of accepted events in $lr$ time windows when Charlie announces the value of $\bold r_j$, and $S_{lr}(r_j)={n_{lr}(r_j)}/{N_{lr}(r_j)}$. In particular, $S_{vv}(r_j)={n_{vv}(r_j)}/{N_{vv}}$;\\
$n_1(r_j)$: the total number of untagged bits when Charlie announces the value of $\bold r_j$;\\
$s_1(r_j)$: the counting rate of untagged bits when Charlie announces the value of $\bold r_j$;\\
$e^{ph}_1(r_j)$: the phase-flip error rate of untagged bits when Charlie announces the value of $\bold r_j$;\\
$\mean{A}$: the expected value of the quantity $A$.

Here we use the 4-intensity decoy-state method ~\cite{zhou2016making,jiang2019unconditional,yu2019sending} to estimate 
the expected value of  ${{s_1}(r_j)}$  first:
\begin{equation}\label{s1rj2}
\begin{split}
\mean{{s_1}(r_j)}= \frac{\mean{{S}_+(r_j)}-\mean{ {S}_-(r_j)}-2(\mu_y^2-\mu_x^2)\mean{  S_{vv}(r_j)}}{2\mu_x\mu_y(\mu_y-\mu_x)},
\end{split}
\end{equation}
where
\begin{equation}\label{eq:joint1}
\begin{split}
\mean{{S}_+(r_j)}=&{\mu^2_ye^{\mu_x}}\mean{S_{vx}(r_j)}+{\mu^2_ye^{\mu_x}}\mean{S_{xv}(r_j)},\\
\mean{ S_-(r_j)}=&{\mu_x^2e^{\mu_y}}\mean{S_{vy(r_j)}}+{\mu_x^2e^{\mu_y}}\mean{S_{yv}(r_j)}.
\end{split}
\end{equation}

In $xx$ time windows, states sent by Alice and Bob are $\ket{\mu_xe^{i\theta_A}}\ket{\mu_xe^{i\theta_B}}$, and \emph{they} publicly announce the phase information  of those states.
If the phase slice $(\theta_A-\theta_B)$ satisfies the condition~\cite{hu2019sending}
\begin{equation}\label{phase}
1-\lvert \cos(\theta_A-\theta_B)\rvert \le \lambda,
\end{equation}
those accepted events in $xx$ time windows will be used to verify the lower bound of the phase-flip error rate of untagged bits $e^{ph}_1$. Here $\lambda$ is a small positive number to be optimized, and the corresponding phase-slice is $\Delta=2\arccos(1-\lambda)$.  We define the total number of instances  with a value of $\bold r_j$ in this phase-slice as $N_{\Delta}(r_j)$, and
\begin{equation}
\begin{split}
N_{\Delta}(r_j)=\frac{\Delta}{\pi}p_x^2NP^2_{r_j}.
\end{split}
\end{equation}

For any accepted event with a value of $\bold r_j$ in $xx$ time windows that satisfies the phase slice condition in Eq.~\eqref{phase}, if $\cos(\theta_A-\theta_B)\ge 0$ and Charlie announces a click of right or $\cos(\theta_A-\theta_B)\le 0$ and Charlie announces a click of left, it indicates an error accepted event. We denote the total number of error accepted events with a value of $\bold r_j$ as $W_{TX}(r_j)$, and the corresponding counting rate of those events is $T_{\Delta}(r_j)$,and 
\begin{equation}
	\begin{split}
T_{\Delta}(r_j)=\frac{W_{TX}(r_j)}{N_{\Delta}(r_j)}
	\end{split}
\end{equation}

 Then the expected value of $n_1(r_j)$ and  $e^{ph}_1(r_j)$ are
\begin{equation}
\begin{split}
\mean{{n}_1(r_j)}=&N2p_zp_vP_{r_j}\mu_ze^{-\mu_z}\mean{{s_1}(r_j)},\\
\mean{e^{ph}_1(r_j)}=&\frac{\mean{ T_{\Delta}(r_j)}-\frac{1}{2}e^{-2\mu_1}\mean{{S}_{vv}(r_j)}}{2\mu_1e^{-2\mu_1}\mean{s_1(r_j)}},\\
\end{split}
\end{equation}
For all $\bold R$ space, the expected value of $n_1$ and $e^{ph}_1$ are
\begin{equation}\label{e1rj}
\begin{split}
\mean{{n}_1}=&\sum_{j=1}^{m}\mean{{n}_1(r_j)},\\
\mean{e^{ph}_1}=&\frac{\sum_{j=1}^{m}\mean{{n}_1(r_j)}\mean{e^{ph}_1(r_j)}}{\mean{n_1}},\\
\end{split}
\end{equation}
In the calculation, we can use Chernoff bound~\cite{jiang2017measurement,chernoff1952measure} to build the relations between real values and their expected values, and finally obtain the bounds of real values of $n_1$ and $e_1^{ph}$. In addition, we can use the joint constraints of statistical fluctuation~\cite{yu2015statistical,jiang2021higher} to reduce the influence of statistical fluctuation.
 
 Here we give a brief review of the Chernoff bound~\cite{jiang2017measurement}. There are $n$ random samples denoted by $\{X_1, X_2,\cdots,X_n\}$, and the value of each $X_i$ equals to $0$ or $1$. Their sum is denoted by $X=\sum_{i=1}^{n}X_i$. Thus the upper bound and lower bound of the expected value of $X$ is
 \begin{equation}\label{chernoff-1}
 	\begin{split}
 	\phi^{L}(X)=\frac{X}{1+\delta_1},\\
 	\phi^{U}(X)=\frac{X}{1-\delta_2},
 	\end{split}
 \end{equation}
 where the value of $\delta_1$ and $\delta_2$ can be solved from the following equations:
 \begin{equation}
 	\begin{split}
 	(\frac{e^{\delta_1}}{(1+\delta_1)^{1+\delta_1}})^{X/(1+\delta_1)}=\frac{\varepsilon}{2},\\
 	(\frac{e^{-\delta_2}}{(1-\delta_2)^{1-\delta_2}})^{X/(1-\delta_2)}=\frac{\varepsilon}{2},
 	\end{split}
 \end{equation}
 with a given failure probability $\varepsilon$.
 
 Also, the Chernoff bound can help us estimate real values $\varphi$ from their expected values $Y$:
 \begin{equation}
 \begin{split}
 \varphi^{U}(Y)=[1+\delta'_1]Y,\\
 \varphi^{L}(Y)=[1-\delta'_2]Y,
 \end{split}
 \end{equation}
 where $\delta'_1$ and $\delta'_2$ can be solved from
 \begin{equation}
 	\begin{split}
 	(\frac{e^{\delta'_1}}{(1+\delta'_1)^{1+\delta'_1}})^Y=\frac{\varepsilon}{2},\\
 	(\frac{e^{\delta'_2}}{(1-\delta'_2)^{1-\delta'_2}})^Y=\frac{\varepsilon}{2}.
 	\end{split}
 \end{equation}

With the help of {joint constraints of statistical fluctuation} method~\cite{yu2015statistical,zhou2016making}, we can reduce the influence of statistical fluctuations in calculating Eq.~\eqref{e1rj}, and the lower bound of $\mean{n_1}$ is
	\begin{equation}
\begin{split}
{\mean{n_1}}^L=&\frac{N2p_vp_z\mu_z e^{-\mu_z}}{2\mu_x\mu_y(\mu_y-\mu_x)}\{\frac{{e^{\mu_x}\mu_y^2}}{Np_vp_x}\phi^L[\sum_{j=1}^{m}(n_{vx}(r_j)+n_{xv}(r_j))]\\ -&\frac{{e^{\mu_y}\mu_x^2}}{Np_vp_y}\phi^U[\sum_{j=1}^{m}({n_{vy}(r_j)}+{{n_{yv}(r_j)}})]\\
-&\frac{{2(\mu_y^2-\mu_x^2)/m}}{Np_vp_v}\phi^U[\sum_{j=1}^{m}({n_{vv}(r_j)})]\}.
	\end{split}
	\end{equation}

Here we set $\{P_{r_j}=1/m$, $j=1,2\cdots,m\}$ ($m$ is the total modes of redundant space). Then, we can obtain the real value of $n_1$: 
 \begin{equation}
 	n_1=\varphi^L({\mean{n_1}}^L).
 \end{equation}
Similarly, the upper bound of $\mean{e^{ph}_1}$ is
\begin{equation}
\begin{split}
\mean{e^{ph}_1}^U=&\frac{p_vp_z\mu_z e^{-\mu_z}\pi m}{p_x^2\mu_xe^{-2\mu_x}\Delta\mean{n_1}^L }\phi^U\{[\sum_{j=1}^{m}{W_{TX}(r_j)}]\}\\
-&\frac{p_z\mu_z e^{-\mu_z}}{2p_v\mu_x\mean{n_1}^Lm}\phi^L\{[\sum_{j=1}^{m}{n_{vv}(r_j)}]\}.\\
\end{split}
\end{equation}
Then, the real value of $e^{ph}_1$ is
\begin{equation}
e^{ph}_1=\varphi^U({n_1}\mean{e^{ph}_1}^U)/{n_1}.
\end{equation}

\section{Formulas for QBER calculations}~\label{appendixB}
We remind the following notations first:\\
 $n_V$: the total number of accepted events in $vv$ time windows;\\
 $n_D$: the total number of accepted events in $zz$ time windows;\\
 $n_C$: the total number of accepted events in both $vz$ time windows and $zv$ time windows.
 
 Mathematically, the number of accepted events in Eq.~\eqref{eq:simulate} can be related to the counting rates by:
 \begin{equation}
 \begin{split}
 n_V=&Np_v^2[\sum_{j=1}^{m}S_{vv}(r_j)],\\
 n_C=&Np_zp_v\{\sum_{j=1}^{m}P_{r_j}[S_{vz}(r_j)+S_{zv}(r_j)]\},\\
 n_D=&Np_z^2[\sum_{j=1}^{m}P_{r_j}^2S_{zz}(r_j)],
 \end{split}
 \end{equation}
 and then, we can calculate $n_t$ and $E_t$ by $n_t=n_V+n_C+n_D$, $E_t=({n_V+n_D})/{n_t}$. 
 
In normal circumstances, these counting rates $S_{lr}(r_j)$ are independent of redundant space, thus we have $S_{lr}(r_1)=S_{lr}(r_2)=\cdots S_{lr}(r_m)\equiv \tilde S_{lr}$, and
 \begin{equation}
 \begin{split}
 n_V=&Np_v^2m\tilde{S}_{vv},\\
 n_C=&Np_zp_v(\tilde{S}_{vz}+\tilde{S}_{zv}),\\
  n_D=&Np_z^2[\sum_{j=1}^{m}P_{r_j}^2]\tilde S_{zz}\geq Np_z^2\tilde S_{zz}/m,
  \end{split}
 \end{equation}
  we can use this because inequality $\sum_{j=1}^{m}P_{r_j}^2\geq 1/m$ always holds mathematically, and the minimum value is obtained when Alice and Bob select elements $\bold r_j$ in set $\{\bold r_1, \bold r_2,\cdots \bold r_m\} (m\geq 2)$ with equal probability.
   With such a choice of parameters, the QBER value $E_t$ can be expressed by
  \begin{equation}\label{eq:QBER14}
  \begin{split}
  E_t=&\frac{p_v^2m\tilde{S}_{vv}+p_z^2\tilde S_{zz}/m}{p_v^2m\tilde{S}_{vv}+p_z^2\tilde S_{zz}/m+p_zp_v(\tilde{S}_{vz}+\tilde{S}_{zv})}\\
  =&\frac{p_v^2m^2\tilde{S}_{vv}+p_z^2\tilde S_{zz}}{p_v^2m^2\tilde{S}_{vv}+p_z^2\tilde S_{zz}+mp_zp_v(\tilde{S}_{vz}+\tilde{S}_{zv})},
  \end{split}
  \end{equation}
  which has a similar form to 
  Eq.~\eqref{QBER-n}.
  This formula shows it quantitatively how the QBER is reduced by our protocol. Because of the first term in the numerator, which is proportional to “m” times of dark count, the QBER is not always simply descending with “m”, though it descends with “m” in the normal cases when the dark-count term is sufficiently small.
  

\end{appendices}

\bibliography{refs}

\end{document}